\documentclass[twocolumn,times,tighten,dvipsnames]{aastex62}

\usepackage{bm}

\newcommand*\Msolarh[0]{h^{-1} \, \mathrm{M_{\odot}}}
\newcommand*\msolarh[0]{h^{-1} \, \mathrm{M_{\odot}}}
\newcommand*\Mpch[0]{h^{-1}\,\mathrm{Mpc}}

\graphicspath{{./}{figures/}}

\shorttitle{Cluster X-ray Masses with \new{ML}}
\shortauthors{Green et al.}

\newcommand{\new}[1]{\textbf{#1}}
\newcommand{\psf}[1]{\textbf{\color{Plum}#1}}
\renewcommand{\new}[1]{{#1}}
\renewcommand{\psf}[1]{#1}
\newcommand{\rev}[1]{#1}

\begin{document}

\title{Using X-Ray Morphological Parameters to Strengthen Galaxy Cluster Mass Estimates via Machine Learning}

\correspondingauthor{Sheridan~B.~Green}
\email{sheridan.green@yale.edu}

\author[0000-0001-9100-6237]{Sheridan~B.~Green}
\altaffiliation{NSF Graduate Research Fellow}
\affiliation{Department of Physics, Yale University, New Haven, CT 06520, USA}

\author[0000-0002-0144-387X]{Michelle Ntampaka}
\affiliation{Center for Astrophysics $|$ Harvard \& Smithsonian, Cambridge, MA 02138, USA}
\affiliation{Harvard Data Science Initiative, Harvard University, Cambridge, MA 02138, USA}

\author[0000-0002-6766-5942]{Daisuke Nagai}
\affiliation{Department of Physics, Yale University, New Haven, CT 06520, USA}
\affiliation{Department of Astronomy, Yale University, New Haven, CT 06520, USA}
\affiliation{Yale Center for Astronomy and Astrophysics, Yale University, New Haven, CT 06520, USA}

\author[0000-0002-3754-2415]{Lorenzo Lovisari}
\affiliation{Center for Astrophysics $|$ Harvard \& Smithsonian, Cambridge, MA 02138, USA}

\author[0000-0003-1750-286X]{Klaus Dolag} 
\affiliation{University Observatory Munich, M{\" u}nchen, Scheinerstrasse 1, D-81679 M{\" u}nchen, Germany}
\affiliation{Max-Planck-Institut f{\" u}r Astrophysik, Karl-Schwarzschild Strasse 1, D-85748 Garching bei M{\" u}nchen, Germany}

\author[0000-0001-7917-3892]{Dominique Eckert}
\affiliation{Max-Planck-Institut f{\" u}r extraterrestrische Physik, Giessenbachstrasse 1, D-85748 Garching, Germany}
\affiliation{Department of Astronomy, University of Geneva, Ch. d'Ecogia 16, 1290 Versoix, Switzerland}

\author[0000-0003-3175-2347]{John A. ZuHone}
\affiliation{Center for Astrophysics $|$ Harvard \& Smithsonian, Cambridge, MA 02138, USA}

\begin{abstract}
We present a machine-learning approach for estimating galaxy cluster masses, trained using both \textit{Chandra} and \textit{eROSITA} mock X-ray observations of 2041 clusters from the \textit{Magneticum} simulations.
We train a random forest (RF) regressor, an ensemble learning method based on decision tree regression, to predict cluster masses using an input feature set.
The feature set uses core-excised X-ray luminosity and a variety of morphological parameters, including surface brightness concentration, smoothness, asymmetry, power ratios, and ellipticity.
The regressor is cross-validated and calibrated on a training sample of 1615 clusters (80\% of sample), and then results are reported as applied to a test sample of 426 clusters (20\% of sample).
This procedure is performed for two different mock observation series in an effort to bracket the potential enhancement in mass predictions that can be made possible by including dynamical state information.
The first series is computed from \new{\psf{idealized \textit{Chandra}-like} mock cluster observations, with high spatial resolution, long exposure time (1\,Ms), and the absence of background}.
The second series is computed from \new{realistic-condition \textit{eROSITA} mocks with lower spatial resolution, short exposures (2 ks), \psf{instrument effects, and background photons modeled}}.
We report a 20\% reduction in the mass estimation scatter when {\it either series} is used in our RF model compared to a standard regression model that only employs core-excised luminosity.
The morphological parameters that hold the highest feature importance are smoothness, asymmetry, and surface brightness concentration.
Hence these parameters, which encode the dynamical state of the cluster, can be used to make more accurate predictions of cluster masses in upcoming surveys, offering a crucial step forward for cosmological analyses.
\end{abstract}


\section{Introduction} \label{sec:intro}
Galaxy clusters are the largest gravitationally bound objects in the universe.
They are rare, with masses $\gtrsim 10^{14}\, M_\odot$, and their abundance is sensitive to the underlying cosmological model. 
Cluster counts can be used to constrain cosmological parameters, provided 
that there is an accurate way to connect the cluster observables (such as X-ray luminosity or temperature) to the underlying dark matter halo mass \citep[for a recent review see][]{2019SSRv..215...25P}. 

Recent cluster-based constraints are in tension with \textit{Planck} cosmic microwave background (CMB) cosmological constraints.   
For example, Sunyaev-Zeldovich \citep[SZ;][]{1972CoASP...4..173S} surveys find fewer massive clusters than would be expected from the \textit{Planck} fiducial cosmology \citep[e.g.,][]{2016A&A...594A..24P}. 
This tension could be explained by a mass bias \textemdash{} a systematic under-estimation of X-ray based cluster mass estimates based on the hydrostatic assumption at the level of $30-45\%$ \citep{2016A&A...594A..24P,2018MNRAS.477.4957B,2019arXiv190407887Z,2019arXiv190707870M}.
However, a significant mass bias remains controversial. 
First, hydrodynamical cosmological simulations predict a hydrostatic mass bias in the range of $15-40$\% \citep[e.g.][]{2006MNRAS.369.2013R,2007ApJ...655...98N, 2013ApJ...777..151L,2014ApJ...782..107N,2016MNRAS.455.2936S, 2016ApJ...827..112B,2017MNRAS.465.3361H} due to non-thermal pressure support provided by bulk and turbulent gas motions \citep[e.g.][]{2009ApJ...705.1129L, 2014ApJ...792...25N, 2015MNRAS.448.1020S} and temperature inhomogeneities in the intracluster medium (ICM) \citep{2014ApJ...791...96R}.  \new{Recent observational results agree that the hydrostatic bias must be small, at least for relaxed systems \citep[e.g.][]{2016MNRAS.457.1522A,2019A&A...621A..40E,  2019A&A...621A..39E, 2019A&A...621A..41G}.}
Second, the hydrostatic mass bias may also arise from the instrument-dependent systematic uncertainties in X-ray temperature measurements \citep{2015A&A...575A..30S,2015MNRAS.448..814I}.
Finally, some cluster- and large-scale structure-based efforts put constraints on cosmological parameters that are consistent with those from the CMB \citep[e.g.,][]{2015MNRAS.446.2205M, 2016ApJ...832...95D, 2017arXiv170801530D} while others are in tension with them \citep[e.g.,][]{2018arXiv181206076H, 2019arXiv190609262J, 2019arXiv190607729N}.
Given the importance of this problem, concerted efforts are underway to calibrate the cluster mass scales using optical weak lensing measurements of background galaxies \citep[e.g.][]{2014MNRAS.439....2V, 2015MNRAS.449..685H, 2016MNRAS.457.1522A, 2019MNRAS.483.2871D} and CMB lensing \citep[e.g.,][]{2019ApJ...872..170R}.

With next-generation observational surveys, such as the \textit{eROSITA} X-ray survey \citep{Merloni2012}, soon to come online, massive data releases that will offer immense cosmological model constraining power are just around the corner.
The \textit{eROSITA} survey is predicted to identify ${\sim}$93,000 galaxy clusters \new{at or above the 50 photon limit} with $M\gtrsim 10^{13.7}\, \Msolarh$ \citep{Pillepich2012,Pillepich2018}.
The product of spectral temperature and gas mass, $Y_X$, is one of the lowest scatter mass proxies \citep{Kravtsov2006}.
However, many of the \textit{eROSITA} observations will be in the regime of low-photon counts, making $T_X$- and $Y_X$-based cluster mass estimates inaccessible \new{\citep{2014A&A...567A..65B}}. 
The core-excised luminosity ($L_{X,\mathrm{ex}}$) is another lower-scatter mass proxy that does not require $T_X$ measurements; excluding the still poorly understood cluster cores ($r\lesssim 0.15\,R_\mathrm{500c}$) reduces the scatter in \rev{the $Y_X$ mass-$L_{X,\mathrm{ex}}$ \citep{Maughan2007, 2009A&A...498..361P} and weak lensing mass-$L_{X,\mathrm{ex}}$ \citep{Mantz2018} relationships, but does so at the expense of drastically reducing the photon statistics.} 

Methods that provide improvements to $L_X$-based mass estimates for these low-photon \textit{eROSITA} clusters could have a steep payoff.
Even in the low-signal regime, there are subtle observable signals that can offer key insights for improving cluster mass estimates.
Measures of cluster morphology, including surface brightness concentration \citep[e.g.,][]{Santos2008}, centroid shift \new{\citep[e.g.,][]{1993ApJ...413..492M}}, and morphological composite parameters \citep[e.g.,][]{Rasia2013}, provide additional information about a cluster's dynamical state \citep{2015MNRAS.449..199M}, which has been shown to influence the scatter in the \rev{mass-$T_X$ relationship of simulated clusters \citep{2008ApJ...685..118V}, the correlated scatter in the relationship between weak lensing mass and integrated SZ Compton parameter $Y_\mathrm{sph}$ \citep[e.g.,][]{2012MNRAS.426.2046A,2012ApJ...754..119M,2016MNRAS.460.3913S},} and the probability that a cluster is observed \citep{2011A&A...526A..79E, 2011A&A...536A...9P, 2017ApJ...846...51L}. 

Modern machine learning (ML) techniques have been shown to reduce error in mass estimates of galaxy clusters.
The techniques that have been developed use cluster dynamics \citep{Ntampaka2015, Ntampaka2016, Ho2019}, X-ray images \citep{Ntampaka2018}, and multiple wavelength summary statistics \citep{Armitage2019, Cohn2019} as input; similar ML techniques have also been applied to less-massive galaxy groups \citep{2019arXiv190202680C, 2019arXiv190701560M}.
These methods hinge on using ML to extract additional information from complex correlations in the mass-observable relationships.
Here, we use ML to take advantage of the complex correlations among morphological parameters, dynamical state, and cluster mass to improve mass estimates.

Our new X-ray cluster mass measurement technique utilizes cluster dynamical state information, encoded in X-ray morphological parameters, to provide improved, lower-scatter mass estimates relative to a mass-luminosity linear regression.
In addition, we demonstrate that this improvement is obtained even in low-photon count \textit{eROSITA} observations, which makes the inclusion of dynamical state information a promising avenue for future cosmological analyses that depend on robust cluster mass estimates.
In Section \ref{sec:hydrosims}, we introduce the \textit{Magneticum} simulations and mock \textit{Chandra} and \textit{eROSITA} X-ray observations of simulated galaxy clusters used in this work.
In Section \ref{sec:params}, we provide an overview of the X-ray morphological parameters employed as features in our models.
In Section \ref{sec:methods}, we describe the preprocessing of the mock catalog data and several regression methods used to build our models.
We summarize the results of our models in Section \ref{sec:results}, followed by our conclusions and proposed follow-up work in Section \ref{sec:conc}.

Throughout this paper, the WMAP7 $\Lambda$CDM cosmology \citep{Komatsu2011} is used: $\Omega_\mathrm{m} = 0.272$, $\Omega_\Lambda = 0.728$, $\Omega_\mathrm{b}=0.046$, $h = 0.704$, $\sigma_8 = 0.809$, and $n_\mathrm{s} = 0.963$.
The base-10 logarithm is denoted by $\log$.
All errors are quoted at the 68\% level.
The majority of this work is performed using the {\sc scikit-learn} \citep{scikit-learn} Python package.

\section{Hydrodynamical Simulations}\label{sec:hydrosims}
\subsection{The Magneticum Simulations}\label{ssec:magneticum}
Our cluster catalog is built from the \textit{Magneticum}\footnote{\href{www.magneticum.org/}{www.magneticum.org/}} \citep{Dolag2015,Dolag2016,Ragagnin2017} suite of cosmological hydrodynamical simulations.
\textit{Magneticum} uses a WMAP7 cosmology \citep{Komatsu2011} with a range of baryonic physics included. For additional details about the simulations and the included baryonic physics, see, e.g., \cite{Biffi2013,Steinborn2015,Teklu2015,Steinborn2016,Bocquet2016,Remus2017}.

We select clusters from the \textit{Magneticum} \texttt{Box2} and \texttt{Box2b} high-resolution simulations, selected for having sufficient resolution and volume to produce a suitable cluster catalog.
\texttt{Box2} has cubic side length of $352\Mpch$ with a dark matter particle resolution of $M_\mathrm{dm}=6.9\times 10^8\, \Msolarh$ and halo catalogs at $z=0.10$, $0.14$, $0.17$, $0.21$, $0.25$, and $0.29$ (as well as higher $z$, but these are not included in our analysis).
\texttt{Box2b} is larger in volume ($640\Mpch$ on a side), has identical mass resolution, and has cluster catalogs at $z=0.25$ and $0.29$.

\begin{figure}
    \centering
    \includegraphics[width=0.45\textwidth]{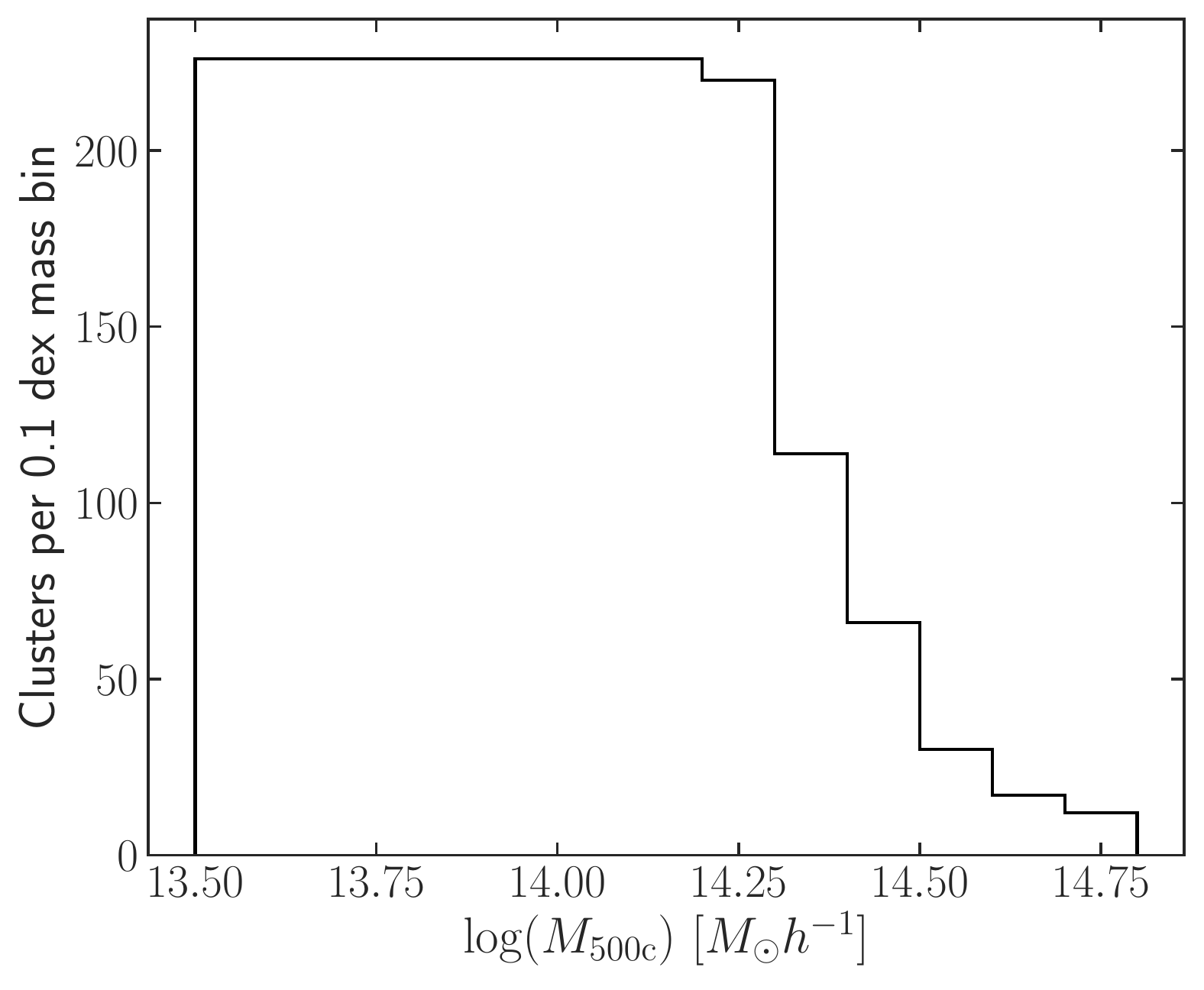}
    \caption{Mass function of the cluster sample used in this work. The sample is flat in the range $10^{13.5} \leq M_\mathrm{500c}/(\msolarh) \leq 10^{14.2}$ and begins decaying in cluster counts for $10^{14.2} \leq M_\mathrm{500c}/(\msolarh) \leq 10^{14.8}$. This sample consists of a total of 2,041 clusters. This uniform distribution in $\log(M_\mathrm{500c})$, our predicted quantity, enables the regression model optimization to equally weight a broad range of cluster masses.}
    \label{fig:mf}
\end{figure}

We initially select all clusters according to their spherical overdensity masses, $M_\mathrm{500c}$,\footnote{We define $M_\mathrm{500c}$ as the mass enclosed within a sphere of (comoving) radius $R_\mathrm{500c}$ whose mean density is 500 times the critical density of the universe at $z=0$.} determined using the \texttt{SUBFIND} algorithm \citep{Springel2001,Dolag2009}.
All clusters above $10^{13.5}\,\msolarh$ are initially included and then subsampled in order to limit the sample to $\leq 230$ clusters per $0.1$\,dex mass bin.
The resulting training catalog has a flat mass function at lower masses, which helps to eliminate mass dependence in the scatter.
Above ${\sim}10^{14.2}\,\Msolarh$, the mass function of this sample falls off, following the mass function of the simulation \citep{Bocquet2016}.
Hence, the sample has a flat mass function in the range $10^{13.5} \leq M_\mathrm{500c}/(\msolarh) \leq 10^{14.2}$ and a falling mass function in the range $10^{14.2} \leq M_\mathrm{500c}/(\msolarh) \leq 10^{14.8}$ \new{(see Fig. \ref{fig:mf})}.
\new{Redshifts in the range $0.1\leq z \leq 0.21$ are roughly equally represented, with ${\sim}$300 clusters per redshift.
However, our sample contains ${\sim}$450 clusters at $z=0.25$ and $z=0.29$ due to the addition of \texttt{Box2b} clusters.}

The final cluster sample includes a total of 2,041 clusters in the redshift range $0.1 \leq z \leq 0.29$, consisting of clusters from both \texttt{Box2} and \texttt{Box2b}.
Within this sample, there are 984 unique clusters, many of which are observed at multiple redshifts.
Based on the assumption of self-similarity \citep{Kravtsov2012}, we verify that \new{the distributions of} all relevant features included in the model exhibit minimal redshift evolution, justifying our inclusion of multiple snapshots for a particular cluster.

\subsection{Mock Observations}\label{ssec:mocks}

From the cluster catalog, we create mock \textit{Chandra} and \textit{eROSITA} observations, employing the \texttt{PHOX} algorithm \citep{2012MNRAS.420.3545B, Biffi2013}.
\new{\texttt{PHOX} models the ICM thermal emission from gas particles by computing the expected number of photons given a fiducial (and large) exposure time and collecting area.
The photon energies are then projected onto the sky plane along a chosen line of sight and cosmologically redshifted.
A foreground galactic absorption model is applied, and models for \textit{Chandra} ACIS-I and \textit{eROSITA} are used to simulate the actual detections.
Further details of the \textit{Magneticum} implementation of \texttt{PHOX} can be found in \cite{2012MNRAS.420.3545B}, \cite{Biffi2013}, and the publicly available \textit{Magneticum} Cosmological Web Portal \citep{Ragagnin2017}.}\footnote{\href{https://c2papcosmosim.uc.lrz.de/}{https://c2papcosmosim.uc.lrz.de/}}

This implementation of \texttt{PHOX} allows the user to select from a number of parameters.
For all observations, we select the ICM-only setting (i.e., AGN are not included as point sources in this work) and employ a 10 Mpc image line-of-sight size to include all relevant correlated structure.
We seek to quantify the level of improvement in cluster mass estimates that can be made possible by incorporating dynamical state information, first in an idealized scenario and then in a realistic case that will be consistent with the observations made in upcoming large, high-throughput surveys such as \textit{eROSITA}.
To this end, our analysis features two different mock observation series: (i) \psf{mocks with \textit{Chandra}-like angular resolution (``idealized \textit{Chandra}'' for short) with a} \textit{Chandra} \new{ACIS-I} instrument area and field of view \new{($2071\times2071$ pixels, 16.9' FoV, 0.49'' pixel)} and a 1 Ms observing time, 
\psf{in the idealized regime of a flat effective area with respect to photon energy ($600\,\mathrm{cm}^2$) and no point spread function (PSF) smearing,} 
as well as (ii) ``realistic \textit{eROSITA}'' observations with an \textit{eROSITA} instrument area and field of view \new{($384\times384$ pixels, $1.03^{\circ}$ FoV, 9.7'' pixel)} with a 2 ks observing time \citep{Merloni2012} \psf{and instrument response and PSF modeled (see \cite{Ragagnin2017} for further details regarding \textit{eROSITA} instrument modeling).}
To more closely imitate the conditions of the upcoming \textit{eROSITA} observations, the ``realistic \textit{eROSITA}'' mock images also include background noise.
The process by which this noise is added is described below.
\new{The \textit{eROSITA} mock observations have a median photon count of ${\sim}$2000 for clusters observed at $z=0.1$ and ${\sim}$100 for clusters observed at $z=0.29$.
In contrast, the \textit{Chandra} mocks have a median photon count of ${\sim}6\times 10^5$ for clusters observed at $z=0.1$ and ${\sim}3\times 10^4$ for clusters observed at $z=0.29$; clearly, derivative quantities computed from the ``idealized \textit{Chandra}'' observations will be affected much less by Poisson noise.}

The cluster bolometric luminosities $L_X$ are calculated by \texttt{PHOX} using the publicly available X-ray package \texttt{XSPEC} \citep{Arnaud1996}.
Core-excised luminosities $L_{X,\mathrm{ex}}$ are computed as follows: (i) compute the total observed photon count $N_\mathrm{tot}$ within $R_\mathrm{500c}$, (ii) compute the observed photon count within $0.15R_\mathrm{500c}$, denoted $N_\mathrm{ce}$, and (iii) scale the bolometric luminosity by the ratio of the photon count observed outside of the core to the total photon count, i.e., $L_{X,\mathrm{ex}} = \frac{N_\mathrm{tot} - N_\mathrm{ce}}{N_\mathrm{tot}} L_X$.
In this work, the core-excised luminosity is used since it has been shown to have lower intrinsic scatter with the cluster mass \citep{Maughan2007,Mantz2018} and is less sensitive to the details of the complicated core physics models used in the simulations.
We note that, for the ``realistic \textit{eROSITA}'' observations, the core-excised photon count ratios are computed \textit{prior to the addition of background noise}.
This likely makes our mock luminosities more accurate than in the case of real \textit{eROSITA} observations.
However, this choice puts the core-excised luminosities from our two observation series on equal footing, such that the model performance differences between the idealized and realistic cases will be dominated by the quality of the morphological parameters.

Redshift is not explicitly included as a feature to train the regression models, however the redshift is used in scaling the luminosity.
Thus, the core-excised luminosity used in this work is always appropriately scaled by the redshift evolution factor, assuming self-similarity, such that we use
\begin{equation}
    L_\mathrm{ex,z} \equiv L_{X,\mathrm{ex}} E(z)^{-7/3} = \frac{N_\mathrm{tot} - N_\mathrm{ce}}{N_\mathrm{tot}} L_X  E(z)^{-7/3} .
\end{equation}

In the $0.5-2.0\,\mathrm{keV}$ energy band, \textit{eROSITA} anticipates an average photon plus particle background of $2.19\times10^{-3}\,\mathrm{counts}\,s^{-1}\,\mathrm{arcmin}^{-2}$ \citep{2018A&A...617A..92C}. 
Thus, for \textit{eROSITA}, the background is given by a Poisson distribution with rate $\lambda=0.113/(2\,\mathrm{ks})$ \citep{Merloni2012}. 
A unique Poisson background is generated for and added to each \textit{eROSITA} mock observation.


\section{Morphological Parameters}\label{sec:params}

In order to encode information about the dynamical state of the cluster into the model, we incorporate various morphological parameters as features, all of which can be directly calculated from the mock X-ray images.
In the following, we define each of these parameters.
We refer the reader to \citet{Lotz2004}, \citet{Rasia2013}, and \citet{2017ApJ...846...51L} for more in-depth discussion on each of the parameters.
Unless otherwise specified, the aperture used to compute the morphological parameters has a radius of $R_\mathrm{ap}=R_\mathrm{500c}$ \new{and is centered on the cluster X-ray peak}; we discuss the implications for this choice at the end of this section.

First, the concentration parameter $c$ quantifies how centrally concentrated the X-ray emission is within the cluster, and has been shown by \citet{Santos2008} to be an indicator for the presence of cooling-core systems at high $z$.
Concentration is defined to be the ratio of the flux within two circular apertures: $0.1R_\mathrm{ap}$ and $R_\mathrm{ap}$ \citep{2017ApJ...846...51L}.

The centroid shift parameter $w$ is defined as the variance of the projected separation between the X-ray peak of the image and the emission centroid obtained within 10 circular apertures of increasing radius up to $R_\mathrm{ap}$ \citep{2017ApJ...846...51L}.

The power ratios, introduced by \citet{Buote1995}, use the ansatz that the X-ray surface brightness profiles are a good tracer of the cluster's projected mass distribution.
The ``power'' is encoded in the coefficients of a 2D multiple decomposition of the cluster X-ray image, where higher-order components probe increasingly smaller scales. The $n$th-order power ratio $P_{n0} = P_n / P_0$ is, in essence, the ratio between the $n$th multipole moments and the $0$th multipole moment.
In this work, we consider $P_{10}$, $P_{20}$, $P_{30}$, and $P_{40}$.
The latter two probe large- and small-scale substructures present within the cluster, and thus further convey dynamical information.

The second power ratio $P_{20}$ provides a measurement of the cluster ellipticity.
Another ellipticity parameter, denoted $e$, is also calculated, defined as the ratio between the semiminor and semimajor axis \citep{2017ApJ...846...51L}.

The asymmetry parameter $A$ quantifies the rotational symmetry of the cluster X-ray emission \citep{Lotz2004}.
$A$ is calculated by rotating by $180^\circ$ and self-subtracting the background-subtracted cluster image from itself, summing the values of the pixels in this image difference and normalizing by the summed pixels in the original image \citep{Abraham1996}.

The smoothness $S$ quantifies the degree of small-scale substructure within the cluster \citep{Lotz2004}. 
$S$ is calculated by boxcar-smoothing and self-subtracting the background-subtracted cluster image from itself, again summing the values of the pixels in this image difference and normalizing by the summed pixels in the original image \citep{Conselice2003}.

Lastly, the $M_{20}$ parameter is an analog of concentration \citep{Lotz2004}.
The total second-order moment of the light is a distance-to-center-weighted sum of the flux $f_i$ within all pixels $i$ in the cluster, $M = \sum_i f_i [(x_i - x_{cc})^2 + (y_i - y_{cc})^2]$, where $cc$ denotes the cluster center.
Then, $M_{20}$ is computed as the ratio of the partial second moment $M_p$, which sums over only the brightest pixels that contain 20\% of the cluster light, divided by the total second moment, written as $M_{20} = \log (M_p / M)$.

These morphological parameters encode dynamical state information.
For example, disturbed clusters tend to be asymmetric (high $A$), clumpy (\textit{high} $S$), and not concentrated (low $c$).
All of the parameters introduced above are calculated for each mock cluster observation, and are used as features in our regression model.

\begin{figure*}
    \centering
    \includegraphics[width=\textwidth]{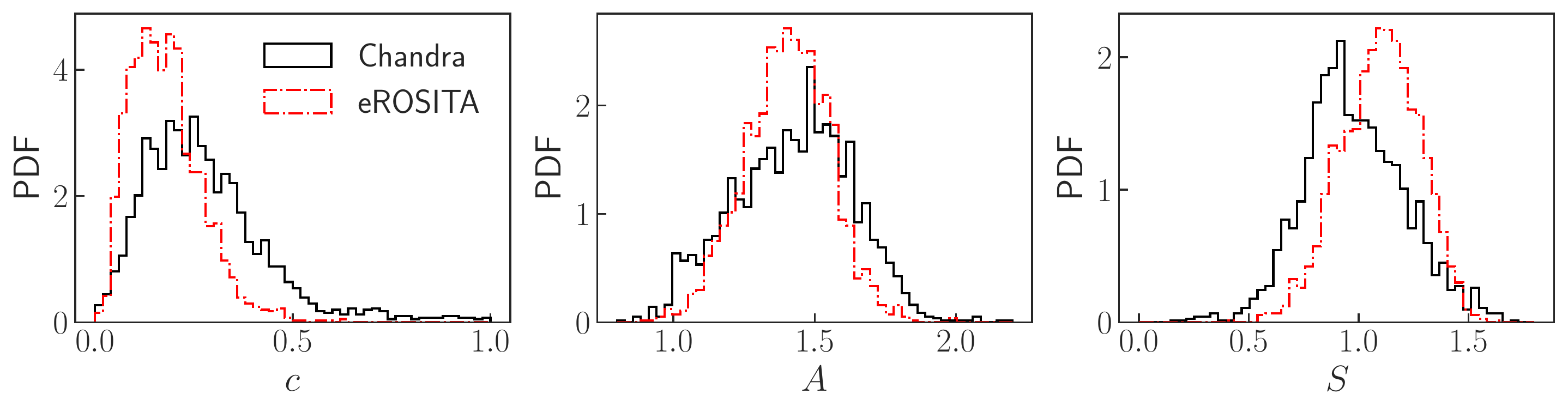}
    \caption{Distributions of the surface brightness concentration $c$, asymmetry $A$, and smoothness $S$ computed from the ``idealized \textit{Chandra}'' and ``realistic \textit{eROSITA}'' mock cluster observation series. Several other morphological parameters are also employed in the analysis (see Sec. \ref{sec:params}), but we find that $c$, $A$, and $S$ are most important for strengthening the cluster mass model.}
    \label{fig:feature_dists}
\end{figure*}

In the subsequent analysis, we utilize two distinct series of morphological parameters, which are computed from our two mock observations series, described above.
The ``idealized \textit{Chandra}'' series is computed from the background-free \textit{Chandra} observations, using $R_\mathrm{ap}=R_\mathrm{500c}$.
The ``realistic \textit{eROSITA}'' series is computed from the \textit{eROSITA} observations with added background, also using $R_\mathrm{ap}=R_\mathrm{500c}$; \new{in this case, the mean background is subtracted prior to computing the parameters}.
The former series is intended to give an upper bound on the expected improvement in cluster mass estimates made possible by including dynamical state information present in idealized, \new{high spatial resolution (0.5'')}, and high-photon count observations.
The latter series is intended to give a more realistic estimate of the expected improvement that will be possible in upcoming cosmological analyses \new{that will be performed with low-photon count cluster observations}.
We acknowledge that by using the exact $R_\mathrm{500c}$ for our aperture  when computing the \new{morphological} parameters, we are neglecting additional scatter that will be present due to this effect. \psf{Also, in the case of the ``idealized \textit{Chandra}'' series, properly including PSF effects would introduce additional smoothing to these observations.} 
Hence, our subsequent results will remain as optimistic estimates.
As we find below, the most important morphological parameters are smoothness, asymmetry, and concentration.
In Fig. \ref{fig:feature_dists}, we plot the distributions of these three parameters, comparing the ``idealized \textit{Chandra}'' and ``realistic \textit{eROSITA}'' series.
While we find generally good agreement between the two series, it is clear that the \textit{eROSITA} cluster observations result in systematically lower concentrations and higher smoothness parameters.


\psf{The lack of high-concentration objects in the \textit{eROSITA} mocks is due to the broader PSF of \textit{eROSITA} with respect to \textit{Chandra}.
Photons originating from the central regions of the observed systems are redistributed over a wider area, which reduces the concentration with respect to the true value.
Since we do not attempt to correct for PSF smearing by applying PSF deconvolution, our procedure for reconstructing $c$ values from \textit{eROSITA} mocks underestimates the concentration of highly-peaked objects.
While also impacted by the broader PSF, the shift to larger $S$ (i.e., \textit{less} smooth) in the \textit{eROSITA} mocks is additionally due to both (i) the lower exposure time, which results in a less ``filled in'' photon distribution due to Poisson noise, and (ii) the presence and subtraction of background, which introduces additional Poisson noise.}

\begin{figure}[h!]
    \centering
    \includegraphics[width=0.23\textwidth]{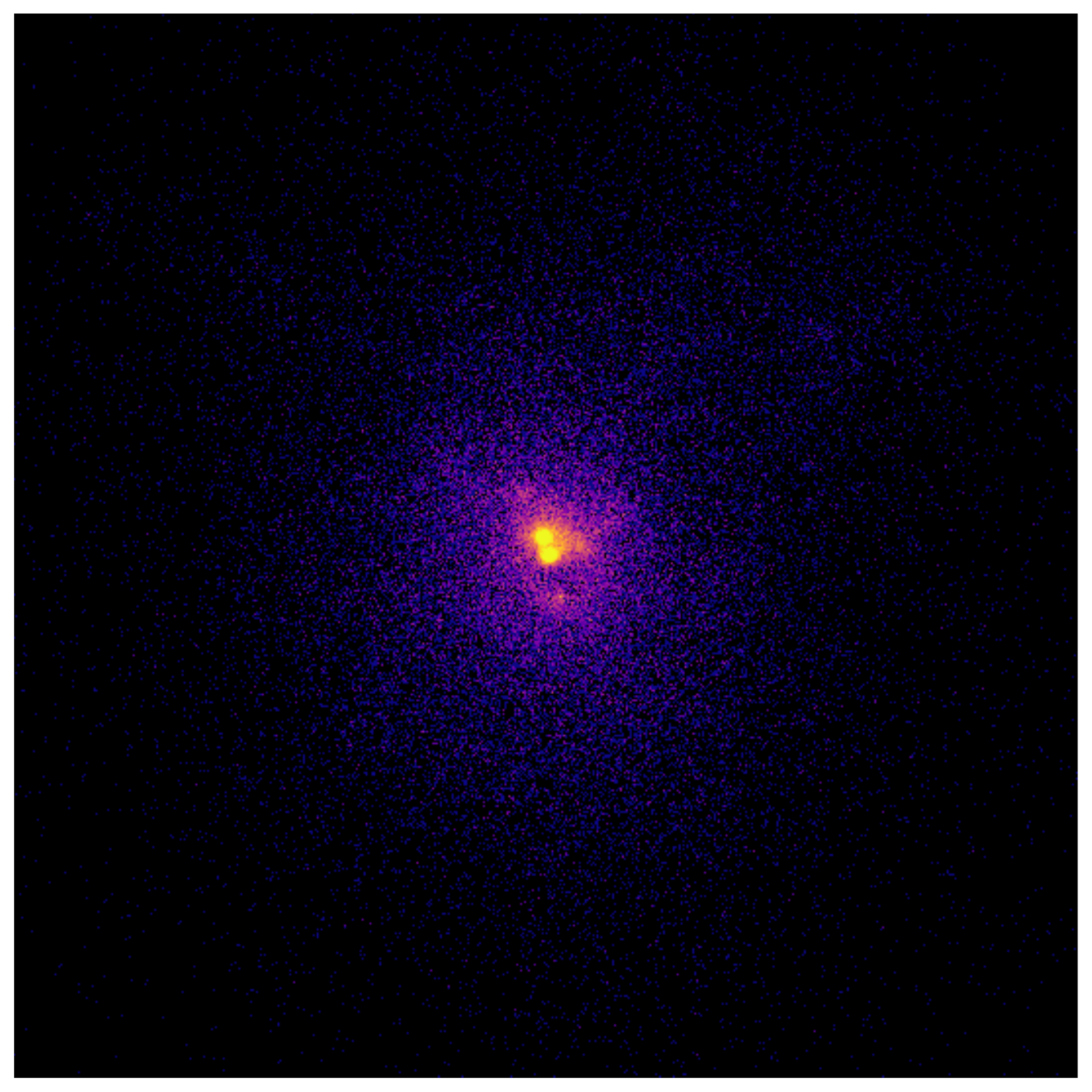}
    \includegraphics[width=0.23\textwidth]{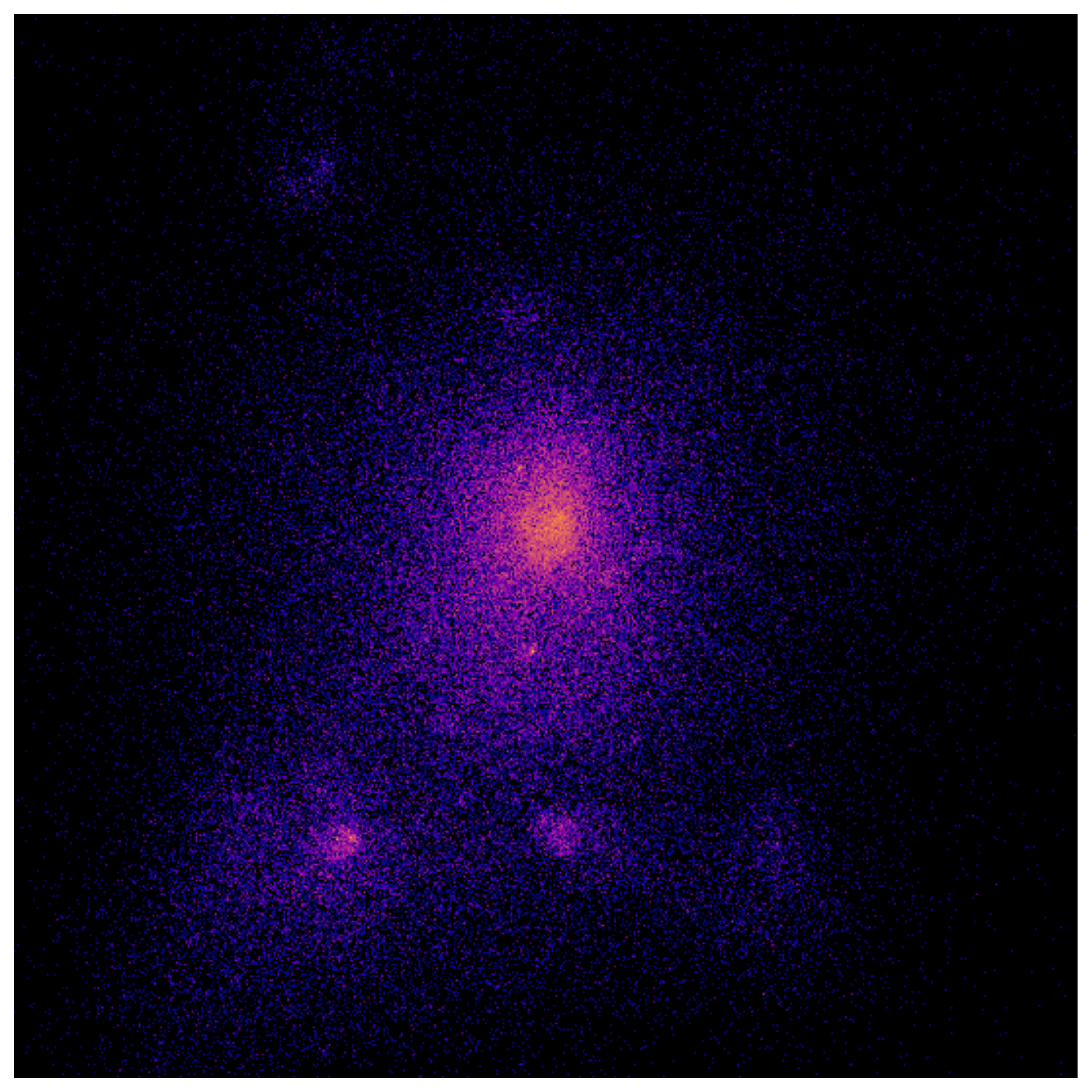}
    \\
    \includegraphics[width=0.23\textwidth]{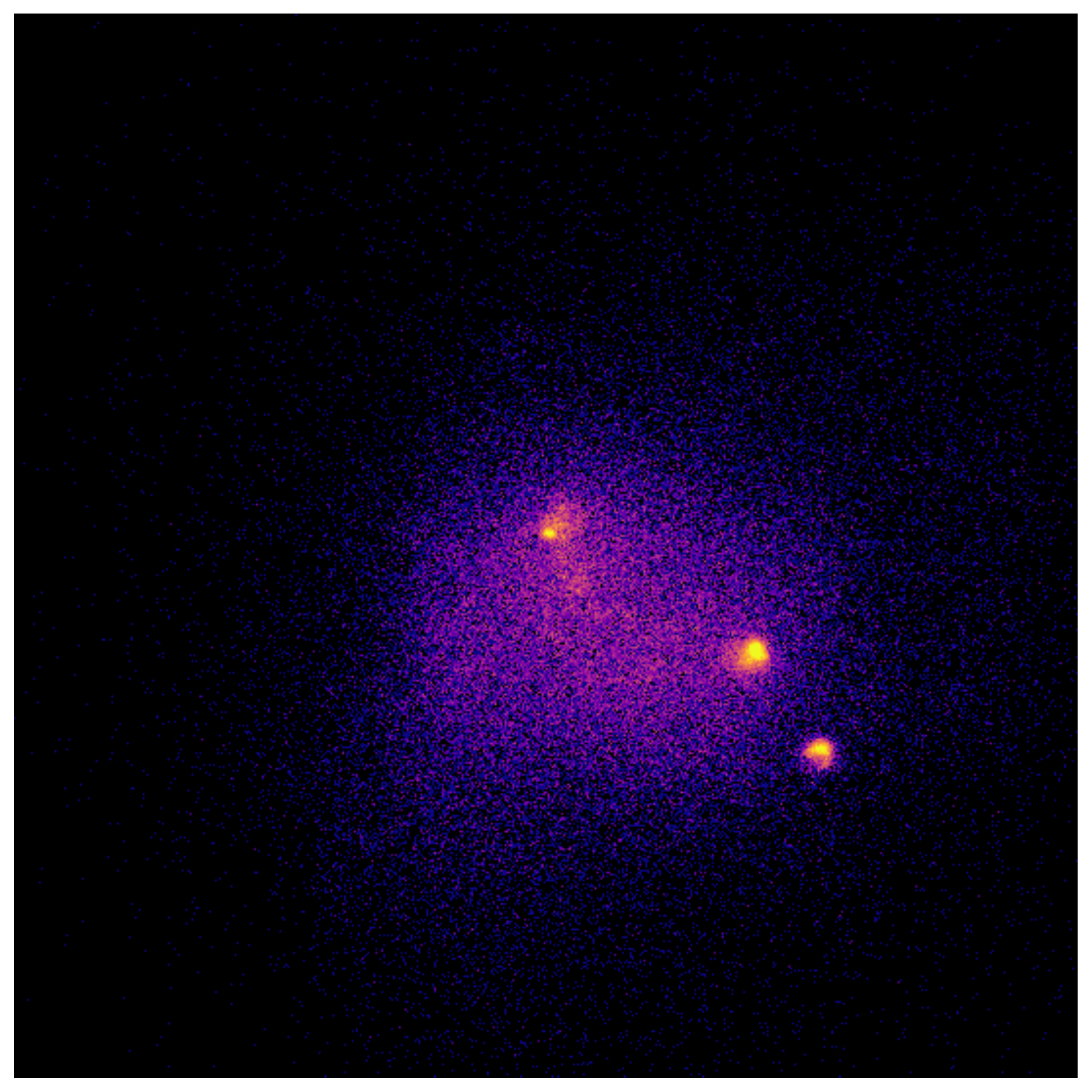}
    \includegraphics[width=0.23\textwidth]{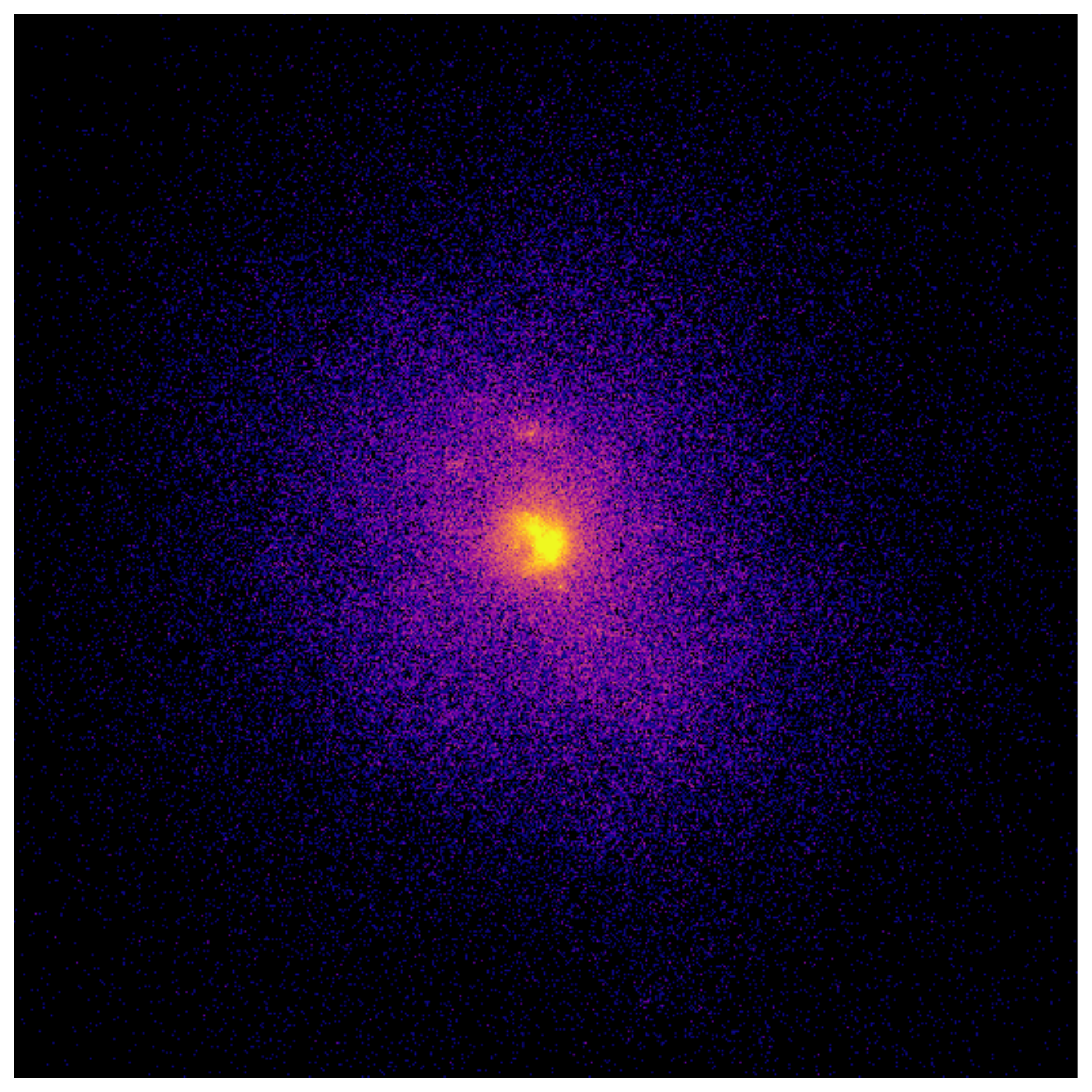}
    \\
    \includegraphics[width=0.23\textwidth]{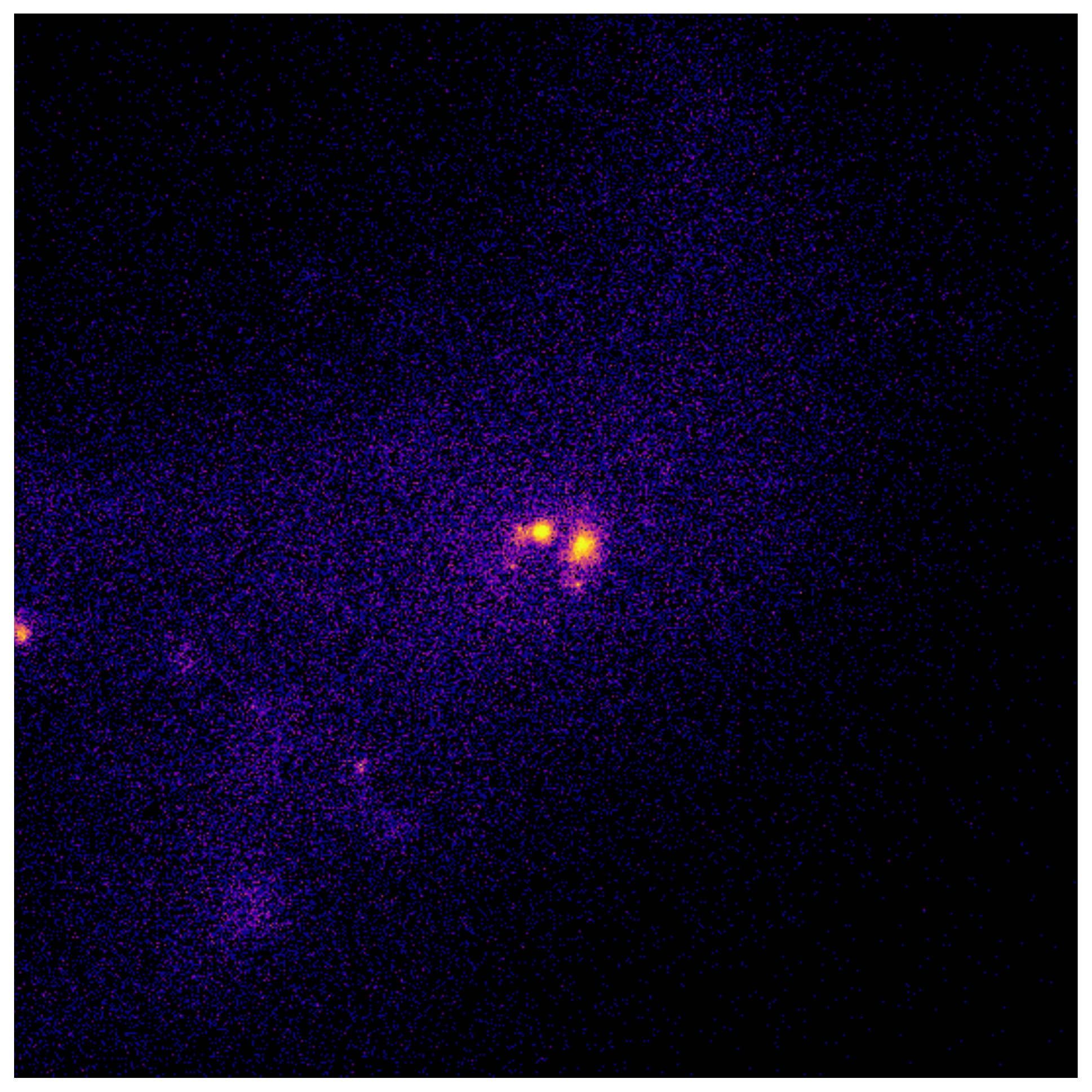}
    \includegraphics[width=0.23\textwidth]{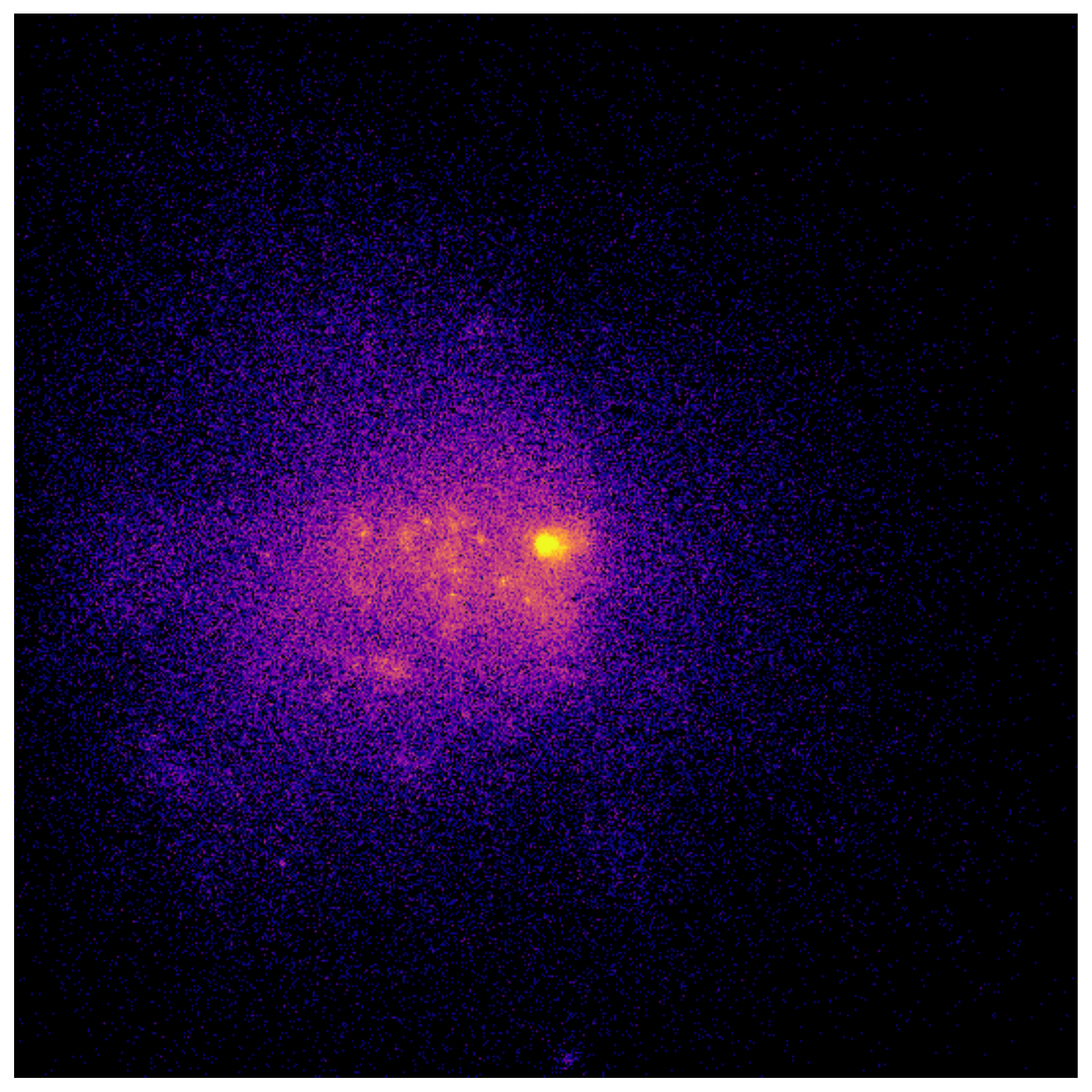}
    \caption{Sample \textit{Chandra}-like cluster images illustrating morphological parameter differences. Each image is centered on the cluster X-ray peak and is cropped a side length of $2R_\mathrm{500c}$. All clusters shown have $10^{14.3} \leq M_\mathrm{500c}/(\msolarh) \leq 10^{14.6}$ at $z=0.1$. Top:  highly concentrated cluster on the left ($c=0.37$) and weakly concentrated cluster on the right ($c=0.04$). Middle: Asymmetric cluster on the left ($A=1.49$) and symmetric cluster on the right ($A=0.93$). Bottom: Less smooth cluster on the left ($S=1.03$) and more smooth cluster on the right ($S=0.61$). Note that a higher value of the smoothness parameter corresponds to a cluster whose surface brightness profile is less smooth. When combined, these parameters, among others (see Sec. \ref{sec:params}), capture the cluster dynamical state by quantifying details such as the presence of substructure or tidal distortions.}
    \label{fig:cluster}
\end{figure}

Additionally, in Fig. \ref{fig:cluster}, we show several example \textit{Chandra} cluster images to demonstrate the morphological parameters, in particular the concentration $c$, asymmetry $A$, and smoothness $S$.
All clusters shown have roughly the same mass, lying in the range $10^{14.3} \leq M_\mathrm{500c}/(\msolarh) \leq 10^{14.6}$, and are all at the lowest redshift of $z=0.1$.
The images are all scaled by the cluster $R_\mathrm{500c}$.
Clearly, a cluster with a larger concentration has a substantially larger fraction of its flux coming from its core.
Furthermore, the asymmetry parameter is successfully able to capture disturbances or substructure in the cluster that result in reduced symmetry.
The smoothness parameter is capable of quantifying small-scale structures; note that a cluster with a larger value of $S$ is overall less smooth, and more likely to contain substructures.

\section{Analysis Methods}\label{sec:methods}
\subsection{Data Preprocessing}\label{ssec:preproc}

As stated previously, our sample consists of 2,041 mock cluster observations across six redshifts in the range $0.1 \leq z \leq 0.29$.
Each observation consists of many features, including core-excised $L_{\mathrm{ex,z}}$ and all of the morphological parameters described in the previous section. 
The logarithm of the power ratios $P_{i0}$, centroid shift $w$, and luminosity are used due to their large dynamic ranges, whereas the remaining features are not transformed.
The regression target for each observation is $\log(M_\mathrm{500c})$ of the cluster.

The sample is then split into a training set that comprises 80\% of the observations (1,617 clusters) and a test set that comprises the remaining 20\% of the observations (425 clusters); \new{this train-test split is a common rule of thumb based on the Pareto principle}.
The split is performed such that all redshift observations of each unique cluster are assigned either to the training or test set, but not split between the two.
For optimization of hyperparameters, $k$-fold cross-validation is employed on the training set, with $k=10$.
The folds are generated such that all observations of each unique cluster are confined to only one fold.

Many regression algorithms require the distribution of each observable to be scaled to have roughly zero mean and unit variance.
In order to scale in such a way that is robust to outliers, we subtract the median and divide by the $1\sigma$ (16th/84th) percentile range computed over the \textit{training} set in order to standardize each feature.
The medians and $\sigma$ are stored from the training set such that an identical transformation is applied to the test set.

\subsection{Regression Methods}\label{ssec:reg}

The work of \citet{Armitage2019} found that ordinary linear regression (OLR) and ridge regression (RR; \citealt{Hoerl1970}) models were able to produce the least scatter in cluster mass estimates using a variety of X-ray, spectroscopic, and photometric datasets.
The authors also tested an ordinary decision tree model \citep{Quinlan1986}, as well as AdaBoost \citep{Freund96} and gradient boosted regression \citep{Friedman2001}, but did not test the popular random forests (RF) regression algorithm \citep{Breiman2001}.
Motivated by their work, and considering that our feature set of morphological parameters contains different information, we will focus our analysis on various linear regression methods and expand by applying random forest regressors.
We train different regression models on our mock catalogs, including a standard mass-luminosity power law ($M-L_{\mathrm{ex,z}}$), an OLR model, several regularized linear regression models (including RR and Lasso regression [LR; \citealt{Tibshirani1996}]), and RF regression models. 

Ordinary linear regression is performed as follows.
For $n$ clusters, each of which are described by $p$ features (i.e., observables), one has a data matrix $\mathbf{X} = \{ \mathbf{x}_0^T, \mathbf{x}_1^T, \mathbf{x}_2^T, ..., \mathbf{x}_n^T\}$, where each $\mathbf{x}$ is a vector of length $p$.
Each vector of observables $\mathbf{x}_i$ is associated with a true logarithmic mass $y_i$.
The mass is predicted as a linear function of the observables, $y_i = \mathbf{x}_i^T \bm{\beta} + \epsilon_i$, where $\bm{\beta}$ is the model parameter vector of length $p$ and $\epsilon_i$ is the random error in the model for cluster $i$.
The best-fit model parameters are chosen by minimizing the cost function, which is selected to be the sum of squared residuals, $E_{\text{OLR}} (\bm{\beta}) = \sum_{i=1}^n (y_i - \mathbf{x}_i^T \bm{\beta})^2$.

In this work, we consider several OLR models.
First, we train a simple OLR model with only one feature, the core-excised luminosity $L_{\mathrm{ex,z}}$.
This allows us to set a baseline for performance and compare the results of our mock observations in terms of the scatter to an observed mass-luminosity relationship.
Then, we train an OLR model on the full feature set, including core-excised luminosity and all morphological parameters.

In an effort to reduce the feature dimensionality and highlight the most important features in the model, one can use regularized linear regression, where an additional term is added to the cost function that introduces a penalty for models with large $\|\bm{\beta}\|$.
In ridge regression, the new cost function is of the form $E_{\text{RR}} (\bm{\beta}; \alpha) = E_{\text{OLR}}(\bm{\beta}) + \frac{\alpha}{2} \|\bm{\beta}\|_2^2$, where $\| \cdot \|_2$ denotes the Euclidean norm.
Similarly, in Lasso regression, the cost function is instead $E_{\text{LR}} (\bm{\beta}; \alpha) = E_{\text{OLR}}(\bm{\beta}) + \alpha \|\bm{\beta}\|_1$, where $\| \cdot \|_1$ denotes the Manhattan norm. 

Regularization acts to reduce the weights of unimportant features in the model, reducing the capability of the model to overfit the training data.
Lasso regression is a more strictly regularized model than ridge regression.
The hyperparameter $\alpha$ is selected via a grid-search cross-validation (CV) of logarithmically-spaced $\alpha$ values, where the model performance is evaluated via $k$-fold CV for each $\alpha$.
In $k$-fold CV, the training set is split into $k$ random subsets (split according to unique cluster ID; see Sec. \ref{ssec:preproc}).
Then, $k-1$ of the subsets are used to train the model, and model predictions are made on the remaining subset.
This process is iterated $k$ times such that predictions are made for all clusters in the training set.
The model performance is quantified by the mean squared error (MSE) of all of the predictions.
The CV process is repeated for all $\alpha$ in the grid, and the $\alpha$ that minimizes the MSE is selected for the training of the final model, which is trained on the full training set and subsequently applied to the test set.

For our last set of models, we use the non-parametric random forest regression model, which is an ensemble technique based on decision trees.
RF models reduce the issues of overfitting that are endemic to decision trees by randomly growing an ensemble of trees, each trained on a different subset of the total training data, and taking the average of their predictions.
Furthermore, RFs increase the tree diversity relative to a standard decision tree ensemble by splitting each node according to the best feature in a random subset of the features, instead of the full feature set.
This increased tree diversity results in a more generalizable model that is less prone to overfitting the training set.
RFs have several important hyperparameters: (i) the number of trees in the forest, (ii) the maximum number of features that can be included in one node splitting condition, (iii) the maximum depth allowed for a tree (i.e., number of decisions that must be made to reach an output), (iv) the minimum number of samples in the training set at a particular node that are required in order for the node to split, (v) the minimum number of samples in the training set required to form a leaf, and (vi) whether or not to use bootstrap resampling (i.e., using ``bagging'' vs. ``pasting'').
Reducing the ``maximum'' hyperparameters (i.e., [ii] and [iii]) or increasing the ``minimum'' hyperparameters (i.e., [iv] and [v]) is an effective way to regularize the model and reduce the tendency for overfitting. The interested reader should refer to \citet{Geron2017} for additional details of various machine learning regressors, including ensemble and tree-based regression.

In this work, we consider \new{several} RF models with different sets of hyperparameters \new{and different input feature sets} in order to demonstrate the level of sensitivity that RF models have to the hyperparameters and to tune an optimal model for future mass predictions.
The first is a RF model with the default hyperparameters from the {\sc scikit-learn} implementation.
The hyperparameters of the second model are optimized using grid search CV over the six-dimensional parameter space of hyperparameters described above.
\new{The third model includes a reduced set of features ($L_{\mathrm{ex,z}}$, $S$, $A$, and $c$), but the hyperparameters are also tuned via grid search CV.}

After selecting hyperparameters for the various models using CV on the training set, the final models are each trained on the entire training set.
The models are then applied to predict the masses of the test set, which we emphasize was never used for either hyperparameter selection or model training, and thus should represent a true example of the generalization capability of the models.
The entire preprocess-split-cross-validate-train-test procedure is performed separately for each of the two series of morphological parameters, i.e., those from the ``idealized \textit{Chandra}'' and ``realistic \textit{eROSITA}'' observations.
In the next section, we report the results for these final models as applied to the test sets.

\section{Results}\label{sec:results}

For both series of observations, we compute the Pearson correlation coefficient between each observable and the cluster mass, shown in Table \ref{tab:corr}.
Additionally, the best fit linear regression model between $\log(M_\mathrm{500c})$ and $\log(L_{\mathrm{ex,z}})$ is used to make mass predictions, and the corresponding mass residuals are then correlated against the observables, also shown in Table \ref{tab:corr}.
The mass residuals $\mathcal{R}$ are defined as
\begin{equation}
    \mathcal{R} = \log(M_\mathrm{500c,pred}) - \log(M_\mathrm{500c,true}),
\end{equation}
where we again emphasize that the base-10 logarithm is used throughout.

\begin{table}[ht!]
\centering
\begin{tabular}{lDDDD}
\tableline
 & \multicolumn{4}{c}{\textit{Chandra}} & \multicolumn{4}{c}{\textit{eROSITA}} \\
 & \multicolumn{4}{c}{Correlation $r$ with} &  \multicolumn{4}{c}{Correlation $r$ with} \\
Feature  & \multicolumn2c{$\log(M_\mathrm{500c})$} & \multicolumn2c{$\mathcal{R}$} & \multicolumn2c{$\log(M_\mathrm{500c})$} & \multicolumn2c{$\mathcal{R}$} \\ \tableline
\decimals
$\log(L_\mathrm{ex,z})$  &  0.927 &  0.000  &  0.929 &  0.000 \\
$c$                       & -0.101 & -0.041  &  0.208 & -0.146 \\
$e$                       &  0.060 & -0.105  &  0.071 & -0.111 \\
$\log(w)$                 & -0.096 &  0.195  & -0.126 &  0.228 \\
$\log(P_{10})$            & -0.170 &  0.212  & -0.197 &  0.224 \\
$\log(P_{20})$            & -0.140 &  0.155  & -0.308 &  0.194 \\
$\log(P_{30})$            & -0.139 &  0.146  & -0.427 &  0.198 \\
$\log(P_{40})$            & -0.145 &  0.149  & -0.489 &  0.207 \\
$A$                       & -0.294 &  0.034  & -0.654 &  0.131 \\
$S$                       & -0.493 &  0.048  & -0.795 &  0.119 \\
$M_{20}$                  &  0.032 &  0.116  & -0.148 &  0.148 \\ \tableline
\end{tabular}
\caption{Pearson correlation between each observable in the model and (i) the true mass, $\log(M_{500c})$, or (ii) the logarithmic mass residual from a mass-luminosity regression, $\mathcal{R}$. These calculations were performed using both the ``idealized \textit{Chandra}'' and ``realistic \textit{eROSITA}'' series of morphological parameters. In both series, $S$ and $A$ correlate most strongly with $\log(M_\mathrm{500c})$. The correlations with mass are generally stronger in the ``realistic \textit{eROSITA}'' series. While $\log(w)$ and $\log(P_{10})$ correlate most strongly with $\mathcal{R}$ in both series, we find that they are not the most important morphological parameters (rather, $S$, $A$, and $c$ are).}\label{tab:corr}
\end{table}

\begin{figure*}
    \centering
    \includegraphics[width=0.45\textwidth]{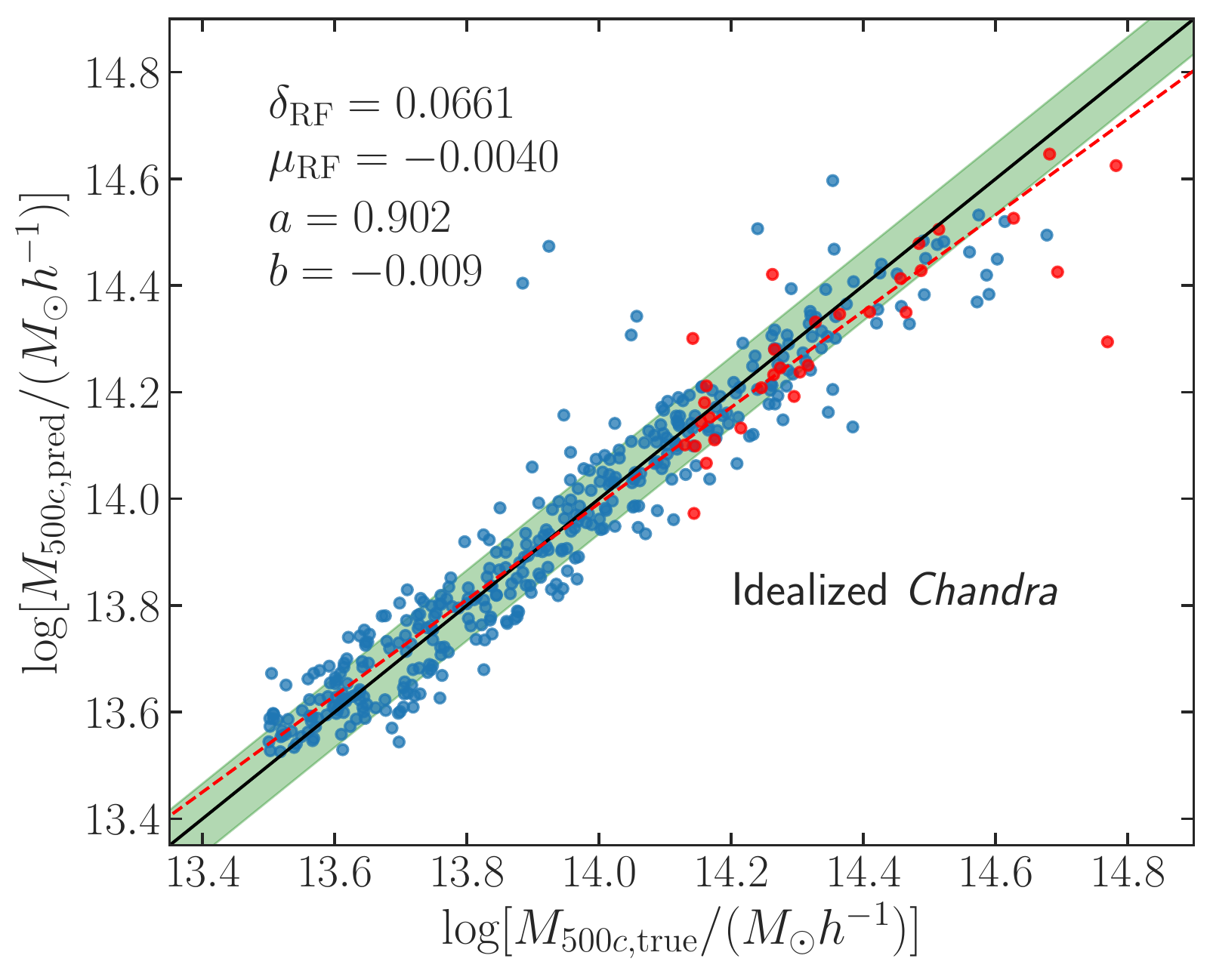}
    \includegraphics[width=0.45\textwidth]{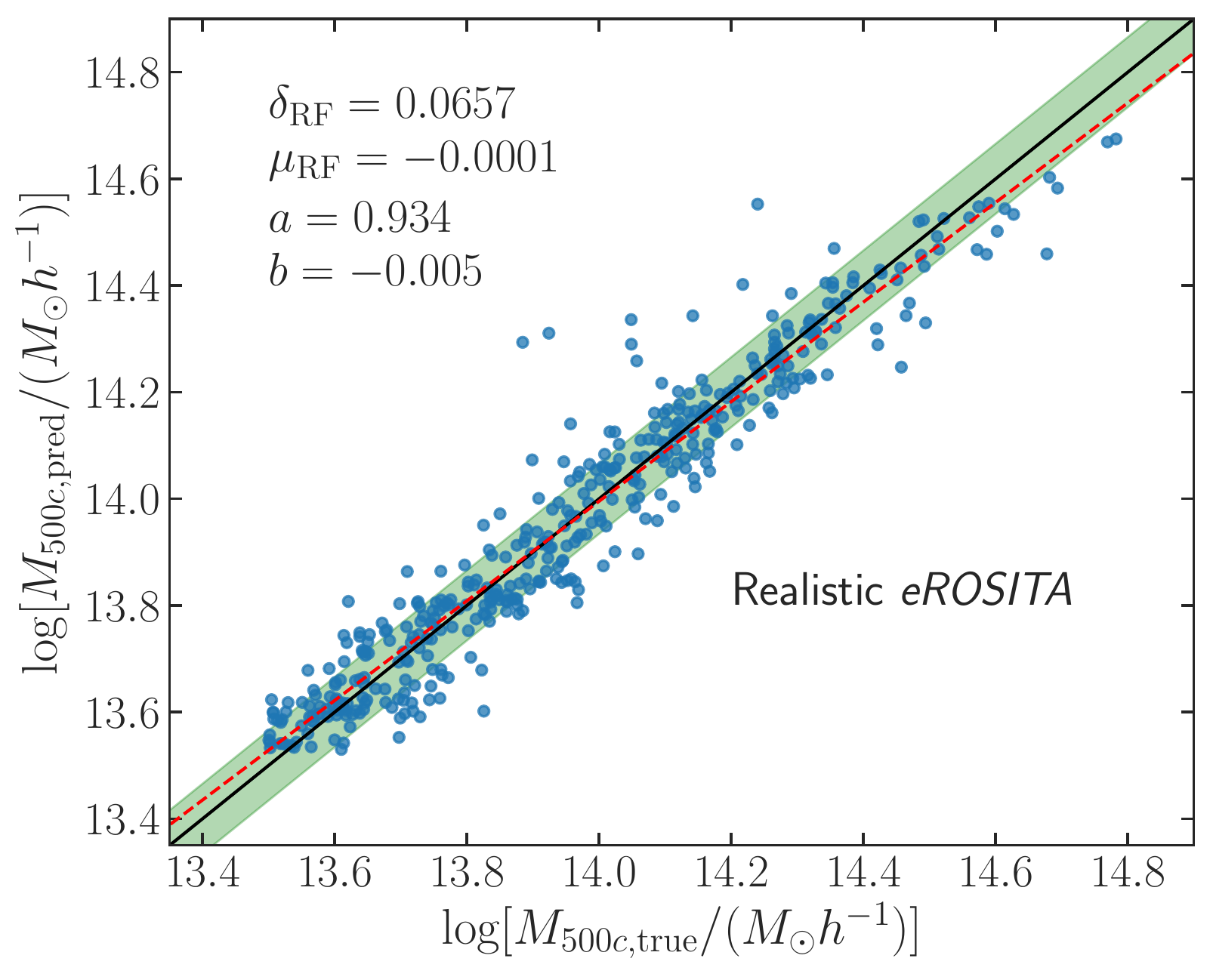}
    \caption{Predicted mass as a function of true mass. Predictions are made using the cross-validation-tuned RF models, which are separately trained using each of the two morphological parameter series. The distributions both have low intrinsic scatter $\delta$ (0.066 dex) and a negligible bias $\mu$. \new{The green band corresponds to the $1\sigma$ scatter, $\delta$.} The dashed red line shows the best power law fit to the predicted masses, $\log (M_{\mathrm{500c,pred}} / [10^{14}\msolarh]) = a \log (M_{\mathrm{500c,true}} / [10^{14}\msolarh]) + b$. For true masses above ${\sim}10^{14.4}\, \Msolarh$, the model consistently underpredicts the mass. This is due to the falling mass function of our sample in the high-mass regime. \new{Additionally, in the case of the \textit{Chandra} observations, some clusters (indicated as red points) extend beyond the instrument field of view, which likely contributes to the lower accuracy of their predicted masses.} Using a training set with a flat mass function that covers the full cluster mass range of interest will likely ameliorate these underpredictions, resulting in a predicted-to-true slope $A$ closer to unity.}
    \label{fig:m_v_m}
\end{figure*}

After luminosity, the observables that correlate or anti-correlate most strongly with mass are smoothness $S$ and asymmetry $A$.
The centroid shift $w$ and first power ratio $P_{10}$ correlate most strongly with the mass residuals, although these correlations are still quite weak ($|r|\approx 0.2$).
Thus, the naive expectation is that $w$ and $P_{10}$ should be the most important additional features (i.e., after luminosity) in a multivariable model of the mass.
However, as we will show below, this ends up not being the case.
We note that while the ranking of the morphological parameters in terms of their correlation strengths remains close to the same between the two mock observation series, the strengths are systematically stronger in the ``realistic \textit{eROSITA}'' observations.
\new{In particular, the high-order power ratios, smoothness, and asymmetry (i.e., the parameters that quantify substructure) correlate much more strongly with mass in the \textit{eROSITA} observations, which is likely a result of the deviation from a smooth profile driven by Poisson noise in the low-photon count regime.}
\new{On the other hand, the correlations between the \textit{Chandra} morphological parameters and the cluster masses are in good qualitative agreement with \citet{2017ApJ...846...51L}, which, using \textit{XMM-Newton} cluster observations, found no significant correlation between the total mass and any of $c$, $w$, or the power ratios.}

The primary model of interest is our cross-validated random forest regressor, which, as we will show below, performs the best among all of the regression methods tested for both series of morphological parameters.
The mass predictions generated by the random forest model for the 426 clusters in the test set are shown in Figure \ref{fig:m_v_m}, with the two separate panels corresponding to the models trained and tested on the two different series of mock observations.
In both cases, it is clear that the model begins to systematically underpredict the masses of the high-mass clusters with $M_\mathrm{500c} \gtrsim 10^{14.4}\, \Msolarh$, which roughly corresponds to the regime where our sample transitions from a flat to falling mass function.
In order to employ this method to predict the masses of observed clusters, it is crucial that the training sample consists of a flat mass function that covers the entire range of masses of interest.
The performance of machine learning models, such as RFs, will greatly improve as larger training samples that are uniform in the prediction (in this case, the mass) become available, for example from state-of-the-art cosmo-hydrodynamical simulations.

\begin{figure}
    \centering
    \includegraphics[width=0.45\textwidth]{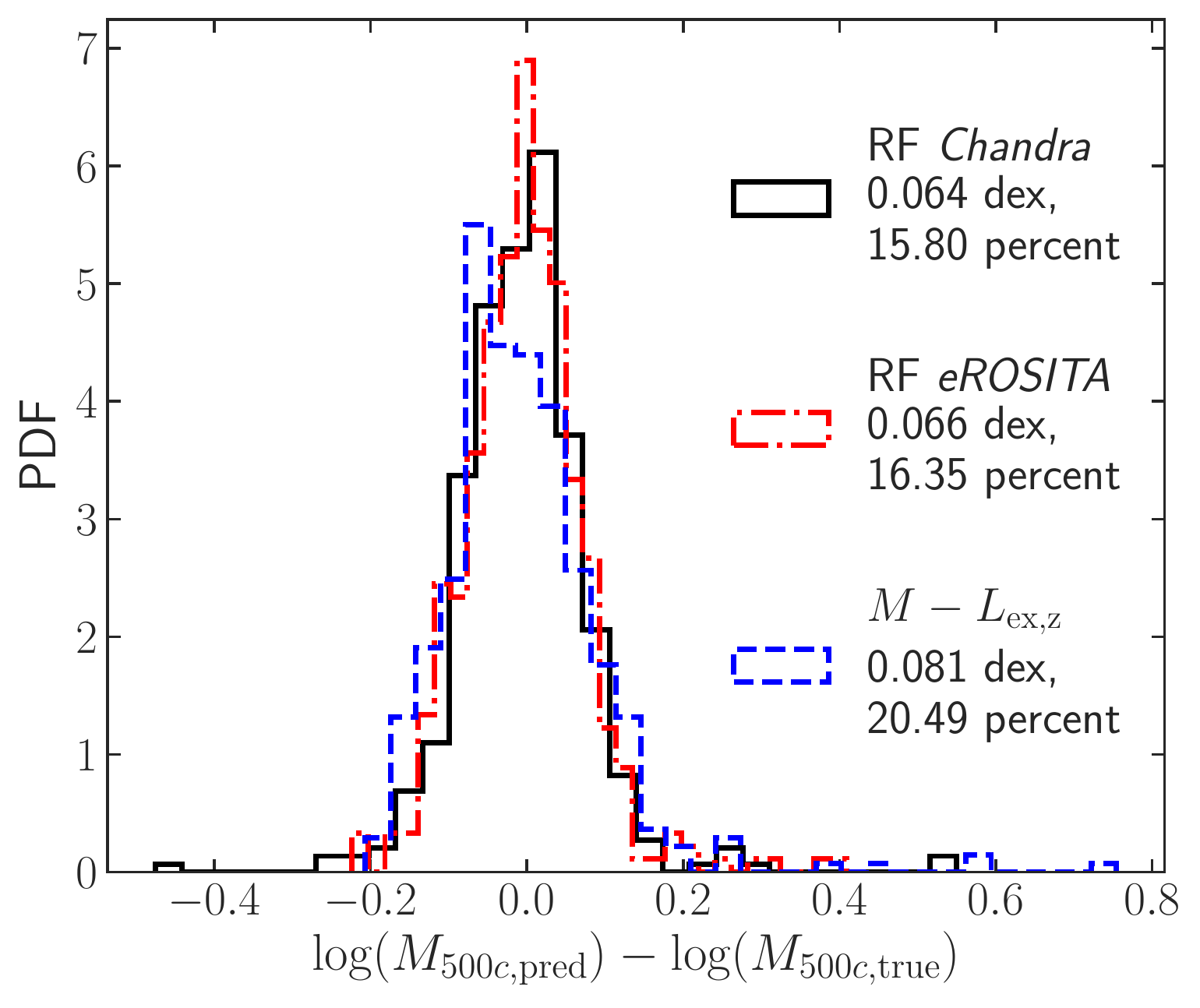}
    \caption{PDF of mass residuals for the cross-validation-tuned RF models of both the ``idealized \textit{Chandra}'' and ``realistic \textit{eROSITA}'' observation series. For comparison, we plot the PDF of mass residuals for the mass-luminosity relationship, with $L_{\mathrm{ex,z}}$ computed using core-excised luminosities from the ``idealized \textit{Chandra}'' observations. In both cases, the RF model offers a ${\sim}20\%$ reduction in scatter relative to the mass-luminosity approach, with negligible bias.}
    \label{fig:errors}
\end{figure}

The PDFs of the mass residuals for these cross-validated random forest regression models are shown in Figure \ref{fig:errors}.
Additionally, the $1\sigma$ intrinsic scatter (i.e., half of the 16th--84th percentile range of $\mathcal{R}$) in the test set for each of the trained models and both of the observation series are shown in Table \ref{tab:scatter}.
For our test sample, the mass residuals of the standard mass-luminosity relationship have a bias of $\mu=-0.017$ dex and $1\sigma$ scatter of $\delta=0.081$ dex.
Interestingly, for both the ``idealized \textit{Chandra}'' and ``realistic \textit{eROSITA}'' observations, the mass residuals have virtually negligible biases and $1\sigma$ intrinsic scatter of $\delta=0.066$ dex, which amounts to a 20\% reduction in scatter relative to the mass-luminosity relationship.
Table \ref{tab:scatter} demonstrates that ordinary linear regression with a combined input feature set that includes the luminosity and all morphological parameters improves only marginally over the single variable mass-luminosity regression.
The incorporation of regularization (i.e., the RR and LR models) does not result in an improved model.
The lack of improvement in these linear models after the inclusion of morphological parameters illustrates that the relationship between cluster morphology and mass is nonlinear and justifies the use of nonlinear approaches, such as a RF regressor.

\begin{table}[h!]
\centering
\begin{tabular}{lDD}
\tableline
\tableline
Method  & \multicolumn2c{$\delta_\mathrm{iC}$} & \multicolumn2c{$\delta_\mathrm{re}$} \\ \tableline
\decimals
$M-L_{\mathrm{ex,z}}$  &  0.081 & 0.081  \\ 
OLR                    &  0.078 & 0.080  \\ 
RR                     &  0.078 & 0.080  \\  
LR                     &  0.079 & 0.081  \\  
RF, defaults           &  0.070 & 0.067  \\  
RF with CV             &  0.066 & 0.066  \\  
\new{RF, only $L_{\mathrm{ex,z}}$}, $c$, $A$, $S$      &  0.070 & 0.071 \\ \tableline
\end{tabular}
\caption{The $1\sigma$ percentile intrinsic scatter, defined as half of the 16th--84th percentile range of the mass residuals $\mathcal{R}$, for each model as trained in the text when applied to the test set, computed using the ``idealized \textit{Chandra}'' and ``realistic \textit{eROSITA}'' mock observation series, denoted $\delta_\mathrm{iC}$ and $\delta_\mathrm{re}$, respectively. The multivariable linear models (OLR, RR, LR) improve only marginally relative to the mass-luminosity relation, with regularization yielding no improvement. The RF models, which capture nonlinear relationships between the input features and the mass, are able to further reduce the scatter beyond any linear approach.}\label{tab:scatter}
\end{table}

\new{In the ``idealized \textit{Chandra}'' observations, 140 of the most massive clusters have $R_\mathrm{500c}$ that extend beyond the instrument field of view.
This results in changes to the morphological parameters calculated for these clusters; for example, the concentrations will systematically increase (although only slightly since the cluster outskirts have the lowest surface brightness) and parameters that quantify substructure ($S$, $P_{30}$, $P_{40}$) may deviate from the correct value if substructures lie outside of the field of view.
We verified that the reported scatters are insensitive to the presence or removal of these clusters from the dataset.
However, this effect, in addition to the dearth of high-mass clusters in the training sample, is likely responsible for the less accurate mass predictions for high-mass clusters (and lower predicted-to-true slope $a$) when using the \textit{Chandra} observation series.
}

The intrinsic scatter for core-excised luminosity-based mass estimates in the observational literature ranges from $\lesssim 15\%$ \rev{for the weak lensing mass-$L_{X,\mathrm{ex}}$ relationship} \citep{Mantz2018} to 16--21\% \rev{for the $Y_X$ mass-$L_X$ relationship} \citep{Maughan2007}. 
While the work of \citet{Mantz2018} finds a lower scatter than we do for our $M-L_{\mathrm{ex,z}}$ relation, we note that they employ a mass cutoff of $M\geq 3\times 10^{14} \, M_\odot$, which results in a much smaller mass range than covered by our dataset.
The scatter in the $M-L_{\mathrm{ex,z}}$ mass residuals for the subset of our clusters with $M_\mathrm{500c}\geq 3\times 10^{14} \, M_\odot$ is 17\%, which is rather close to that of \citet{Mantz2018}.
The sample used by \citet{Maughan2007} includes clusters down to $8 \times 10^{13}\, M_\odot$, which is more consistent with our cluster sample.
Thus, the scatter in the $M-L_{\mathrm{ex,z}}$ mass residuals of our full sample, $\delta=0.081$ dex (20.5\% scatter), is consistent with similar such calculations performed \rev{using observations of either the $Y_X$ mass-$L_{X,\mathrm{ex}}$ or weak lensing mass-$L_{X,\mathrm{ex}}$ relationships}.
The current state-of-the-art mass estimation methods require high-resolution, long-exposure cluster observations with good spatial and spectral resolution, with the $Y_X$ approach resulting in 5--7\% scatter \citep{Kravtsov2006}.
The observational conditions necessary for utilizing the $M-Y_X$ method will simply not be present for the vast majority of clusters observed in upcoming surveys such as \textit{eROSITA}.
However, we have demonstrated that our method achieves a 20\% improvement in cluster mass estimates over $M-L_{\mathrm{ex,z}}$ even in the low-spatial resolution, short-exposure (2 ks) conditions of \textit{eROSITA} observations ($\delta=0.066$ dex, 16\%).
Since we also find the same level of improvement for our \textit{Chandra} observations, which likely places an upper bound on the method performance, this suggests that the mass-encoding dynamical state information, as quantified by our set of morphological parameters, remains present even in the short-exposure \textit{eROSITA} observations.

One metric available for interpreting the results of the random forest model is the \textit{feature importance} ranking.
For example, in an OLR model, the feature importances are roughly quantified by the magnitudes of the regression coefficients.
The standard metric for RF feature importance, and the one that is implemented in {\sc scikit-learn}, is the \textit{mean decrease in impurity}, which measures how effective a feature is at reducing the variance of the predictions.
Based on this importance measurement, the most important features and their \textit{Chandra} importances are (after $L_{\mathrm{ex,z}}$, 75\%), in decreasing order: smoothness $S$ (10\%), asymmetry $A$ (5\%), and concentration $c$ (4\%).
The ranking of these features are the same for both observation series, with importance magnitudes being similar, and all other morphological parameters have negligible importance ($\lesssim 1\%$) in both cases \new{(see Table \ref{tab:importance})}.

\begin{table}[h!]
\centering
\begin{tabular}{lDD}
\tableline
\tableline
 & \multicolumn{4}{c}{Importance (\%)} \\
Feature  & \multicolumn2c{\textit{Chandra}} & \multicolumn2c{\textit{eROSITA}} \\ \tableline
\decimals
$L_{\mathrm{ex,z}}$     &  74.9 & 76.2  \\ 
$S$                     &  10.0 & 14.6  \\ 
$A$                     &  4.8  &  2.5  \\  
$c$                     &  3.9  &  1.5  \\ \tableline
\end{tabular}
\caption{\new{The feature importances for the cross-validated RF models, computed based on the mean decrease in impurity. The remaining morphological parameters are omitted from the table, as their importances are all $\lesssim 1\%$. The smoothness $S$, asymmetry $A$, and surface brightness concentration $c$ encode the most additional mass information after the core-excised luminosity $L_{\mathrm{ex,z}}$.}}\label{tab:importance}
\end{table}

Another measurement of feature importance, known as \textit{permutation importance}, quantifies the drop in the $R^2$ score when the values of a feature are permuted over the samples.
Thus, the larger the drop in $R^2$ when a feature is permuted, the more important the feature.
This metric is considered less biased, as it is not sensitive to the dynamic range of the input variables; this detail is likely irrelevant since our features are scaled.
When this importance metric is employed, we find the same feature importance ranking as before for the ``realistic \textit{eROSITA}'' parameter series.
However, for the ``idealized \textit{Chandra}'' series, we find that, following luminosity, $c$ and $S$ are nearly tied for being most important, followed by $A$.
These three features are not the highest correlators with mass residual, which goes against the naive expectation that the most important features after luminosity should correlate the strongest with mass residual.
However, $S$ and $A$ do correlate the most strongly with mass after $L_{\mathrm{ex,z}}$, as seen in Table \ref{tab:corr}.
Thus, there must be some nonlinear relationship between $L_{\mathrm{ex,z}}$, $S$, $A$, and $c$ (slightly supplemented by the combination of all the other morphological parameters) that the RF model identifies in order to make the improved mass estimates.

\new{Some of the morphological parameters employed, particularly $w$ and the power ratios, will be more difficult to measure accurately for \textit{eROSITA}-observed clusters.
Motivated by our finding that $S$, $A$, and $c$ are the most important morphological parameters, we consider an additional RF model that is cross-validated and trained with a reduced feature set that includes only these three parameters and $L_{\mathrm{ex,z}}$.
As displayed in Table \ref{tab:scatter}, we find that this reduced model yields a $1\sigma$ scatter of $\delta=0.070$ for the ``idealized \textit{Chandra}'' series and $\delta=0.071$ for the ``realistic \textit{eROSITA}'' series.
While this is still a 12--13\% improvement over the mass-luminosity relation, this finding highlights the benefit of including the additional morphological parameters, even considering that each of them has an importance of $\lesssim 1\%$.
The combined effect of the additional morphological parameters, including $e$, $w$, $M_{20}$, and the power ratios, is ultimately responsible for roughly a third of the overall improvement offered by our approach.
}

\section{Conclusion}\label{sec:conc}
In this work, we have presented a method for estimating cluster masses from mock X-ray observations of galaxy clusters.
The mock observations are generated from 2,041 clusters with masses in the range of $10^{13.5} \leq M_\mathrm{500c}/(\msolarh) \leq 10^{14.8}$ and over a redshift range of $0.1 \leq z \leq 0.29$ from the \texttt{Box2} and \texttt{Box2b} \textit{Magneticum} simulations.
The mass predictor is based on a random forest regression model, trained with a feature set that includes only the core-excised luminosity and a set of morphological parameters, all of which can be computed directly from an X-ray image.

We demonstrate that this method can be used to estimate the masses of galaxy clusters with negligible bias and a scatter of $\delta=0.066$ dex (16\%).
This model achieves a 20\% reduction in the scatter relative to a more standard core-excised luminosity power law.
Importantly, the same level of improvement is present both when using idealized, high-resolution, long-exposure (1 Ms) \textit{Chandra} mock observations with no background and when using realistic, low-resolution, short-exposure (2 ks) \textit{eROSITA} mock observations with added background noise.
\new{The majority of this improvement comes from three parameters: smoothness $S$, asymmetry $A$, and surface brightness concentration $c$.}
\new{A more conservative model, which includes only the luminosity and these three parameters, estimates the cluster masses with a scatter of $\delta=0.070$, demonstrating that a third of the overall improvement comes from the inclusion of the additional morphological parameters ($e$, $w$, $M_{20}$, and the power ratios).}
However, it is yet to be seen how additional sources of error present in real observations will affect the performance of this model; for example, the scatter in $R_\mathrm{500c}$ measurements will propagate to increased scatter in the morphological parameters.

While excising the cluster core reduces the scatter in mass estimates, this improvement comes at the cost of lowering the photon counts used in the analysis.
\new{However, even at the \textit{eROSITA} detection threshold of ${\sim}$30 core-excised photon counts, the statistical uncertainty on $L_{\mathrm{ex,z}}$ will be $\lesssim 20\%$; hence, the mass estimate errors from a mass-luminosity relationship will be dominated by intrinsic scatter even in the low-photon limit.}
Since the dynamical state of the cluster encodes important information that affects mass errors, the inclusion of morphological parameters, which utilize the full photon distribution, in the mass model enable more accurate predictions \new{with reduced intrinsic scatter relative to $M-L_{\mathrm{ex,z}}$}.
\new{However, we expect that the statistical uncertainty in the mass estimates will still closely follow the corresponding statistical uncertainty of $L_{\mathrm{ex,z}}$.}
The relationship between the morphological parameters, the luminosity, and the mass is complicated and nonlinear; we have demonstrated that the nonlinear RF regression method offers a substantial improvement over linear models.

\rev{Our model was trained to predict the spherical overdensity masses, $M_\mathrm{500c}$, of galaxy clusters identified in \textit{Magneticum} by the \texttt{SUBFIND} algorithm.
As demonstrated in the \citet{Knebe2011} halo finder comparison project, this algorithm is able to estimate the $M_\mathrm{200c}$ masses of NFW host haloes to within $\lesssim 3\%$; more broadly, all modern halo finders compared in \citet{Knebe2011} are able to determine host halo $M_\mathrm{200c}$ to within $\lesssim 10\%$.
Hence, we expect that uncertainty introduced due to the mass estimates of our simulated clusters is sub-dominant, but not insignificant, compared to the intrinsic scatter of the $M-L_{\mathrm{ex,z}}$ relationship.
The recent \textit{FABLE} simulations project \citep{Henden2018}, a set of hydrodynamical simulations with similar sub-grid physics to \textit{Magneticum}, reports that the $M-L_{X}$ of simulated clusters is in excellent agreement with the observed relation based on X-ray hydrostatic masses.
Thus, if there is indeed an X-ray hydrostatic mass bias, this would indicate that simulated clusters may be too gas-rich, resulting in $L_{X}$ values that are high relative to weak lensing-calibrated $M-L_{X}$.
In addition, current models of AGN feedback result in simulated cluster cores that do not match the observed cool-core and non-cool-core cluster populations.
Because of this, the morphological parameters of observed clusters that depend most sensitively on the cluster core (i.e., X-ray surface brightness concentration) may be biased relative to observations.
We expect that these simulation sources of uncertainty will improve as the hydrostatic mass bias quandary approaches a resolution and as more sophisticated AGN feedback models are developed.
}

Random forest models are notoriously bad at extrapolation.
We expect that our model's systematic underprediction of the masses of clusters in the high-mass tail of the halo mass function will improve when trained on a cluster sample that is uniform across the full mass range of interest.
When large-volume, high-resolution hydrodynamical simulations become available for creating such a training sample, we expect that this can be used to train a model that predicts well across the entire mass range.
This model, once trained on a sufficiently large simulated sample, could then be applied to a set of \textit{Chandra}-observed clusters, such as the HIFLUGCS sample \citep{Zhang2011}, and the predictions could be tested against accurate mass estimates, such as those based on $Y_X$.

ML-based methods of estimating galaxy cluster masses from X-ray observations, including the method presented here as well as others in the literature \citep[e.g.,][]{Ntampaka2018}, offer a promising step towards extracting the maximum information content present in imminent datasets such as \textit{eROSITA}.  Modern ML methods will enable the completion of an unprecedentedly accurate cosmic census and position the halo mass function to be used to place ever stronger cosmological constraints.
Ultimately, the continued progress in cosmological hydrodynamical simulations, both in terms of physical realism and size, are rapidly facilitating the coming of age of these techniques, which will soon be ready for deployment on state-of-the-art cluster observation samples.

\acknowledgments
The authors would like to thank Stefano Ettori for providing the original morphological parameter estimation code.
SBG is supported by the US National Science Foundation Graduate Research Fellowship under Grant No. DGE-1752134.
LL acknowledges support from NASA through contracts 80NSSCK0582 and 80NSSC19K0116.
KD acknowledges the support of the DFG Cluster of Excellence `Origins'.

\software{\texttt{SCIKIT-LEARN} \citep{scikit-learn}, \texttt{XSPEC} \citep{Arnaud1996}}

\bibliography{bibliography}

\end{document}